\documentclass[aps,pra,
reprint,
 amsmath,amssymb,amsfonts,
 showpacs]{revtex4-1}
\usepackage{bm}
\usepackage{graphicx}
\begin{document}
\title{Space-time description of pulsed laser field by means of coherent states}
\author{Adam Bechler}
\affiliation{Institute of Physics, University of Szczecin, Wielkopolska 15, 70-451 Szczecin, Poland}
%\ead{adamb@univ.szczecin.pl}

\begin{abstract}
Space-time description of pulsed laser radiation by means of coherent states is presented. The corresponding displacement operator contains space dependent annihilation and creation operators, not the "standard" operators corresponding to mode decomposition of the electromagnetic field. This allowed for a direct description of space and time dependence of the electromagnetic field of the pulse. The main characteristics of the pulse is its profile in the wave-vector space. It determines space-time structure of expectation values of the electromagnetic field carried by the pulse, and the distribution of photon modes as well. Determination of the latter required transition from the space-time, or field description to the photon mode approach. Expectation values of the electric field of a Gaussian pulse propagating in the $z$-direction and linearly polarized in the $x$-direction are presented graphically in the form of contour plots in the $x-z$ plane. Apart of the dominant $x$-component electric field has a small longitudinal component in the $z$-direction. The dominant component of the magnetic field in the $y$-direction is acompannied by small components in $z$- and $x$-directions. The plane wave limit of the pulse can be implemented by shrinking transverse width of  the profile to zero (transverse spatial extension goes to infinity), while keeping the photon density constant. One can then retrieve standard plane wave form of the pulse also in the dipole approximation.
\end{abstract}
\maketitle
\section{Introduction}
Laser radiation in the form of strong and short pulses finds nowadays application in many branches of physical research. The most popular example is the interaction of strong and short pulses with atomic and molecular systems, and in particular ionization of atoms and molecules. For strong fields with maximum intensities not exceeding one atomic unit ($\sim 10^{16}\,\mathrm{W/cm^2}$) dipole approximation is sufficient to account for major features of atomic and molecular ionization ~\cite{Milosevic}. In such cases kinetic energy of the ejected electron is small compared to its rest energy, so that motion of the photoelectron can be described in the framework of non-relativistic dynamics in a plane-wave electromagnetic field \cite{DiChiara}. However, the dipole and plane wave approximation fails in the case of ultrastrong tightly focused pulses, when spatial dependence of the field becomes important ~\cite{DiChiara} and non-relativistic description of the electron motion becomes inadequate.
In the purely classical approach description of focused laser radiation can be based on various forms of three dimensional non-paraxial approximation ~\cite{Lax,Davis}. Solutions of Maxwell equations of this type were recently applied to the description of Xe-photoionization in ultrastrong fields ~\cite{DiChiara} characterized by the values of parameters for which the standard one-dimensional dipole approximation is not justified.

With the advent of ultrastrong short pulses of radiation for which the dipole approximation does not provide a sufficient theoretical description, it is desirable to formulate a theoretical approach accounting both for the propagation effects and the transverse structure of the pulse. One of the possibilities is given by non-paraxial description mentioned above, which is based on approximate solution of Maxwell equations in classical electrodynamics without taking the quantum effects in electromagnetic field into account. The purpose of this paper is to propose fully quantum description of spatial and temporal dependence of an ultrastrong pulse of radiation by means of coherent states. Although to a good approximation laser radiation can be described by classical solutions of Maxwell equations, its quantum nature is also important. In particular, the use of coherent states allows not only to analyze space images of the pulse field (as in the purely classical approach), but also to determine directly distribution of photon modes in the pulse. This is not possible in the case of a purely classical soluion, where the photons are simply "absent", and, among various states of quantum electromagnetic field the coherent state gives a good approximation to the real laser field~\cite{Huayu}. Coherent states have been also used for the description of light beams in a recent publication ~\cite{IBB3} dealing with the uncertainty relation for light beams.   

Electromagnetic field of a pulse of radiation is localized in a bounded region of space moving with the speed of light in the direction of pulse propagation. Exact classical solutions of Maxwell equations describing focused or otherwise bounded configurations of electromagnetic field have been discussed in the past ~\cite{Hellwarth,Ziolkowski,IBB1,IBB2}. On the other hand, it has been shown by I. Bia\l ynicki-Birula and Z. Bia\l ynicka-Birula that, in general, the photons "cannot be sharply localized" ~\cite{IBB}, with the only exception given by coherent states, when full localization of photons is possible. In general, quantum limitations on the localization of energy density do not apply to quantum states having classical counterparts ~\cite{IBB1}, and coherent state is a well known example. For these reasons coherent states of quantum electromagnetic field offer, as will be seen below, an adequate description of pulsed laser radiation.

The main idea of the present paper is to construct the coherent state using space-dependent annihilation and creation operators, introduced for the first time by Glauber ~\cite{Glauber1, Glauber2} and applied recently~\cite{IBB} in the discussion of photon localization problem.  The advantage of using space-dependent operators instead of the standard monochromatic mode operators is that the approach of this type allows to obtain in a simple and direct way space and time profiles of the electric and magnetic fields of the pulse beyond dipole approximation. On the other hand, an important question is in what way one can determine the distribution of photons inside the pulse. The answer to this question requires transition from the space-time description to the, complementary to it, photon mode description, or, in other words, transition from the field to the photon picture of the pulse. It can be also considered as the question of transition from the wave-packet modes to the monochromatic modes of quantum electromagnetic field, discussed in the paper by Smith and Raymer~\cite{Smith}. This approach can be also used to find the photon-mode structure of the pulse of radiation.

The main characteristics of the coherent state describing the pulse of radiation is its profile $\boldsymbol{\varphi}(\mathbf{r})$ in the configuration space ~\cite{IBB} and its Fourier transform in the wave-vector space $\widetilde{\boldsymbol{\varphi}}(\mathbf{k})$. The latter will be assumed in the Gaussian form both in the longitudinal direction (direction of propagation) and in the transverse direction. The width of the profile in the wave-vector space in the longitudinal direction will be assumed larger than in the transverse plane of the pulse, which corresponds to a pulse with transverse dimension in the configuration space larger than in the longitudinal direction. These assumptions about the Gaussian profile of the pulse allowed for the approximate analytic calculation of the expectation values of electric and magnetic field in the coherent state describing pulsed laser radiation. Expectation values of the fields were further analyzed as functions of coordinates and time, and represented graphically in the form of contour plots in the $z-x$ plane, where $z$ is the direction of propagation of the pulse and $x$ is the direction of linear polarization. In general, the electric field has two components; the dominant $x$-component and the longitudinal component in the direction of propagation, so that the pulse does not constitute a purely transverse electromagnetic wave, although the longitudinal part is a few orders of magnitude smaller than the transverse one. The dominant component of the magnetic field is in the $y$-direction and is accompanied by small components in the $z$- and $x$-directions. The non-transverse contributions to electromagnetic field of the pulse vanish after sufficiently long time, when the pulse approaches a linearly polarized plane wave.

The paper is organized as follows. In Section 2, which has a review character, I give a brief summary of the formalism of time and coordinate dependent annihilation and creation operators based on~\cite{IBB}. Section 3 contains expressions for the expectation values of the vector potential and discussion of the photon modes of the coherent state describing the pulse. The space-time structure of the pulse is discussed in details in Section 4, which also contains graphical representations of the space images of the electric and magnetic field. Section 5 contains final remarks and some details of the calculations have been moved to Appendices.    
\section{Space-dependent annihilation and creation operators and coherent states}
In this review part I briefly summarize for the sake of completeness the formalism of space dependent annihilation and creation operators, presented in Ref~\cite{IBB}. The operators of quantum electromagnetic field have the following decomposition into the annihilation and creation parts, analogous to the positive and negative frequency decomposition
\begin{subequations}\label{eq:1}
	\begin{equation}\label{eq:1a}
		\hat{\mathbf{D}}(\mathbf{r},t)=\left(\frac{\hbar c\varepsilon_0}{2}\right)^{1/2}
		\left[\hat{\mathbf{d}}^\dagger(\mathbf{r},t)+\hat{\mathbf{d}}(\mathbf{r},t)\right],
	\end{equation}
	\begin{equation}\label{eq:1b}
		\hat{\mathbf{B}}(\mathbf{r},t)=\left(\frac{\hbar c\mu_0}{2}\right)^{1/2}
				\left[\hat{\mathbf{b}}^\dagger(\mathbf{r},t)+\hat{\mathbf{b}}(\mathbf{r},t)\right],
	\end{equation}
\end{subequations}
where the electric annihilation and creation operators are denoted by $\hat{\mathbf{d}}$ and $\hat{\mathbf{d}}^\dagger$ respectively, and the corresponding magnetic operators by $\hat{\mathbf{b}}$ and $\hat{\mathbf{b}}^\dagger$. The electric and magnetic operators are not independent and fulfill the relations
\begin{equation}\label{eq:2}
	\hat{\mathbf{b}}(\mathbf{r},t)=-i\hat{\chi}\hat{\mathbf{d}}(\mathbf{r},t),\quad
	\hat{\mathbf{b}}^\dagger(\mathbf{r},t)=i\hat{\chi}\hat{\mathbf{d}}^\dagger(\mathbf{r},t),
\end{equation}
where $\hat{\chi}$ is the helicity operator which in the wave-vector space acts as $\hat{\chi}=(i/k)\mathbf{k}\times$,
and in the position space its action on a vector function $\mathbf{f}(\mathbf{r})$ looks as ~\cite{IBB}
\begin{equation}\label{eq:3}
	\hat{\chi}\mathbf{f}(\mathbf{r})=\frac{1}{2\pi^2}\int d^3r'\frac{1}{|\mathbf{r}-\mathbf{r}'|^2}\nabla\times\mathbf{f}(\mathbf{r}').
\end{equation}
The equal time commutation relations between components of the $\mathbf{d},\,\mathbf{d}^\dagger$, and $\mathbf{b},\,\mathbf{b}^\dagger$ operators have a nonlocal character
\begin{subequations}\label{eq:4}
	\begin{equation}\label{eq:4a}
		[d_i(\mathbf{r},t),d_j^\dagger(\mathbf{r}',t)]=\left(\partial_i\partial_j-\delta_{ij}\Delta\right)\frac{1}{2\pi^2|\mathbf{r}-\mathbf{r}'|^2},
	\end{equation}
	\begin{equation}\label{eq:4b}
			[b_i(\mathbf{r},t),b_j^\dagger(\mathbf{r}',t)]=\left(\partial_i\partial_j-\delta_{ij}\Delta\right)\frac{1}{2\pi^2|\mathbf{r}-\mathbf{r}'|^2},
		\end{equation}
\end{subequations}
whereas the commutator of $\mathbf{b}$ and $\mathbf{d}^\dagger$ is local
\begin{equation}\label{eq:5}
	[b_i(\mathbf{r},t),d_j^\dagger(\mathbf{r}',t)]=-i\epsilon_{ikj}\partial_k\delta^{(3)}(\mathbf{r}-\mathbf{r}').
\end{equation}
The remaining commutators vanish.\\
Due to nonlocal character of the commutation relations \eqref{eq:4} the Fock basis costructed with the use of either the electric or magnetic creation operators are not orthogonal, and both basis are mutually orthogonal due to the local commutator \eqref{eq:5} \cite{IBB}. Coherent state is defined with the use of space-dependent annihilation and creation operators as
\begin{equation}\label{eq:6}
	|\varphi\rangle=\exp\int d^3r \left[\bm{\varphi}(\mathbf{r})\cdot\mathbf{d}^\dagger(\mathbf{r})-
	\bm{\varphi}^*(\mathbf{r})\cdot\mathbf{d}(\mathbf{r})\right]|0\rangle,
\end{equation}
where $|0\rangle$ is the vacuum state and the space profile of the coherent state $\bm{\varphi}(\mathbf{r})$ is given by a divergenceless complex vector function. Coherent state is the eigenstate of annihilation operator
\begin{equation}\label{eq:7}
	\mathbf{d}(\mathbf{r})|\varphi\rangle=\int\frac{d^3k}{(2\pi)^3}|\mathbf{k}|
	\widetilde{\bm{\varphi}}(\mathbf{k})e^{i\mathbf{k}\cdot\mathbf{r}}|\varphi\rangle,
\end{equation}
where $\widetilde{\bm{\varphi}}(\mathbf{k})$ is the Fourier transform of the space profile of the coherent state,
\begin{equation}\label{eq:8}
	\widetilde{\bm{\varphi}}(\mathbf{k})=\int d^3re^{-i\mathbf{k}\cdot\mathbf{r}}\bm{\varphi}(\mathbf{r}).
\end{equation}
Eq.\eqref{eq:8} can be easily obtained from the definition of coherent state, using the commutation relation \eqref{eq:4a}. In the similar way
\begin{equation}\label{eq;9}
 \mathbf{b}(\mathbf{r})|\varphi\rangle=\int\frac{d^3k}{(2\pi)^3}\mathbf{k}\times
 \widetilde{\bm{\varphi}}(\mathbf{k})e^{i\mathbf{k}\cdot\mathbf{r}}|\varphi\rangle.
\end{equation}
\section{Electromagnetic field of the pulse and photon modes}
This section is devoted to the description of electric and magnetic field of the coherent pulse, and also to the discussion of its photon modes. Electromagnetic field of the pulse is given by expectation values,
\begin{subequations}\label{eq:10}
	\begin{equation}\label{eq:10a}
		\bm{\mathcal{D}}(\mathbf{r},t)=\langle\varphi|\hat{\mathbf{D}}(\mathbf{r},t)|\varphi\rangle,
	\end{equation}
	\begin{equation}\label{eq:10b}
		\bm{\mathcal{B}}(\mathbf{r},t)=\langle\varphi|\hat{\mathbf{B}}(\mathbf{r},t)|\varphi\rangle.
	\end{equation}
\end{subequations}
To find the expectation values at an arbitrary time it is necessary to know action of the annihilation operator $\hat{\mathbf{d}}(\mathbf{r},t)$ on the coherent state. Using the Heisenberg equations of motion,
\begin{equation}\label{eq:11}
	\frac{\partial \hat{\mathbf{d}}(\mathbf{r},t)}{\partial t}=\frac{i}{\hbar}\left[\hat{\mathcal{H}},\hat{\mathbf{d}}(\mathbf{r},t)\right]
\end{equation}
with the Hamiltonian
\begin{equation}\label{eq:12}
	\mathcal{H}=\frac{\hbar c}{2}\int d^3r
	[\hat{\mathbf{d}}^\dagger(\mathbf{r},t)\hat{\mathbf{d}}(\mathbf{r},t)+
	\hat{\mathbf{b}}^\dagger(\mathbf{r},t)\hat{\mathbf{b}}(\mathbf{r},t)],
\end{equation}
one obtains with the use of Eqs.\eqref{eq:4}
\begin{equation}\label{eq:13}
	\hat{\mathbf{d}}(\mathbf{r},t)=\int d^3r'\int\frac{d^3k}{(2\pi)^3}e^{-i\mathbf{k}\cdot\mathbf{r}'}e^{-ikx}\hat{\mathbf{d}}(\mathbf{r}'),
\end{equation}
where $kx$ denotes scalar product of the four-vectors $k$ and $x$ in the Minkowski space, and
\begin{equation}\label{eq:14}
	k=(|\mathbf{k}|,\mathbf{k}),\quad x=(ct,\mathbf{x}).
\end{equation}
Substitution of \eqref{eq:7} into \eqref{eq:13} gives
\begin{equation}\label{eq:15}
	\hat{\mathbf{d}}(\mathbf{r},t)|\varphi\rangle=
	\int\frac{d^3k}{(2\pi)^3}|\mathbf{k}|\widetilde{\bm{\varphi}}(\mathbf{k})e^{-ikx}|\varphi\rangle,
\end{equation}
and the corresponding formula for the magnetic annihilation operator reads
\begin{equation}\label{eq:16}
	\hat{\mathbf{b}}(\mathbf{r},t)|\varphi\rangle=
	\int\frac{d^3k}{(2\pi)^3}\mathbf{k}\times\widetilde{\bm{\varphi}}(\mathbf{k})e^{-ikx}|\varphi\rangle.
\end{equation}
Using the decomposition \eqref{eq:1} of the field operators into space dependent annihilation and creation operators and the formulae \eqref{eq:15} and \eqref{eq:16} it is straightworward to calculate the expectation values \eqref{eq:10},
\begin{equation}\label{eq:17}
	\bm{\mathcal{D}}(\mathbf{r},t)=-\varepsilon_0\partial_t\bm{\mathcal{A}}(\mathbf{r},t),\quad
	\bm{\mathcal{B}}(\mathbf{r},t)=\nabla\times\bm{\mathcal{A}}(\mathbf{r},t),
\end{equation}
where the vector potential has the form
\begin{equation}\label{eq:18}
	\bm{\mathcal{A}}(\mathbf{r},t)=-i\sqrt{\frac{\hbar}{2\varepsilon_0c}}\int\frac{d^3k}{(2\pi)^3}
	[\widetilde{\bm{\varphi}}(\mathbf{k})e^{-ikx}-
	c.c.].
\end{equation}
Obviously, a result of this type should be expected, since, in principle, any solution of the wave equation can be written in this form. However, by using the decomposition \eqref{eq:1} and the relations \eqref{eq:15} and \eqref{eq:16} one can retrieve the correct overall factor and also determine correctly distribution of photons in the pulse. Solution of classical wave equation without quantum attributes of the coherent state does not give such possibilities. At most, the photon mode distribution can be in this case determined only in a heuristic way.

To determine the photon mode distribution of the pulse it is necessary to "translate" the space-time description of the coherent state into the photon mode description. The standard annihilation and creation operators of the monochromatic photon modes are normalized to fulfill the commutation relation
\begin{equation}\label{eq:19}
	[\hat{a}^{(\lambda)}_\mathbf{k},\hat{a}^{(\lambda ')\dagger}_\mathbf{k'}]
	=2(2\pi)^3|\mathbf{k}|\delta^{(3)}(\mathbf{k}-\mathbf{k}')\delta_{\lambda\lambda '},
\end{equation}
where $\lambda$ is the polarization index. With this normalization the decomposition of the electric field operator into positive and negative frequency parts has the form ~\cite{Smith}
\begin{equation}\label{eq:20}
	\hat{\mathbf{D}}(\mathbf{r},t)=i\sqrt{\frac{\hbar\varepsilon_0}{c}}\sum_{\lambda}
	\int d\Gamma_\mathbf{k}\omega_\mathbf{k}\hat{a}^{(\lambda)}_\mathbf{k}\mathbf{e}^{(\lambda)}_\mathbf{k}e^{-ikx}+h.c.,
\end{equation}
where $d\Gamma_\mathbf{k}=[2(2\pi)^3|\mathbf{k}|]^{-1}d^3k$ is the invariant measure on the light cone in the reciprocal space and $\mathbf{e}^{(\lambda)}_\mathbf{k}$ is the polarization vector. Annihilation operator of the wave packet mode is given by ~\cite{QED}
\begin{equation}\label{eq:21}
	b_\alpha=\sum_\lambda\int d\Gamma_\mathbf{k}f^{(\lambda)*}_\alpha(\mathbf{k})a^{(\lambda)}_\mathbf{k},
\end{equation}
where the wave packet profiles fulfill the relations of completness and orthogonality
\begin{subequations}\label{eq:22}
	\begin{equation}\label{eq:22a}
		\sum_\alpha f^{(\lambda)*}_\alpha(\mathbf{k})f^{(\lambda ')}_\alpha(\mathbf{k}')=
		2(2\pi)^3|\mathbf{k}|\delta^{(3)}(\mathbf{k}-\mathbf{k}')\delta_{\lambda\lambda'}
	\end{equation}
	\begin{equation}\label{eq:22b}
		\sum_\lambda\int d\Gamma_\mathbf{k}f^{(\lambda)}_\alpha(\mathbf{k})f^{(\lambda)*}_\beta(\mathbf{k})=\delta_{\alpha\beta},
	\end{equation}
\end{subequations}
which implies the usual commutation relation for the annihillation and creation operators of the wave packet modes
\begin{equation}\label{eq:23}
	[b_\alpha,b^\dagger_\beta]=\delta_{\alpha\beta}.
\end{equation}
The formula inverse to \eqref{eq:21} has the form
\begin{equation}\label{eq:24}
	a^{(\lambda)}_{\mathbf{k}}=\sum_{\alpha}f^{(\lambda)}_\alpha(\mathbf{k})b_\alpha.
\end{equation}
The coherent state (\ref{eq:6}) can be written with the use of the wave packet annihillation and creation operators (\ref{eq:21}) as
\begin{equation}\label{eq:25}
	|\varphi\rangle=\exp\sum_{\alpha}(A_\alpha\hat{b}^\dagger_\alpha-A_\alpha^*\hat{b}_\alpha)|0\rangle,
\end{equation}
where the quantities $A_\alpha$ are expressed by the coherent state profiles $\bm{\varphi}$ and the wave packet profiles $f^{(\lambda)}_\alpha$ as
\begin{equation}\label{eq:26}
	A_\alpha=-\frac{i}{\sqrt{2}}\sum_{\lambda}\int\frac{d^3k}{(2\pi)^3}\widetilde{\bm{\varphi}}(\mathbf{k})
	\cdot\mathbf{e}^{(\lambda)*}_\mathbf{k}f^{(\lambda)*}_\alpha(\mathbf{k}).
\end{equation}
Details of the derivation of Eqs. (\ref{eq:25}) and  (\ref{eq:26}) are given in the Appendix \ref{A}.

The photon number operator is given by $\hat{N}=\sum_{\alpha}\hat{b}^\dagger_\alpha\hat{b}_\alpha$ and its average value in the coherent state $|\varphi\rangle$, $\langle\hat{N}\rangle_\varphi=\sum_{\alpha}|A_\alpha|^2$, can be also expressed explicitly by the profile of the coherent state
\begin{equation}\label{eq:27}
		\langle\hat{N}\rangle_\varphi=\sum_{\lambda}\int\frac{d^3k}{(2\pi)^3}|\mathbf{k}|
		|\widetilde{\mathbf{\varphi}}(\mathbf{k})\cdot\mathbf{e}^{(\lambda)*}_\mathbf{k}|^2.
\end{equation}
It follows immediately from (\ref{eq:27}) that the distribution of polarized photons in the wave vector space is given by
\begin{equation}\label{eq:28}
	dN^{(\lambda)}_\mathbf{k}=|\mathbf{k}|
			|\widetilde{\mathbf{\varphi}}(\mathbf{k})\cdot\mathbf{e}^{(\lambda)*}_\mathbf{k}|^2
			\frac{d^3k}{(2\pi)^3},
\end{equation}
and in the case of unpolarized photons
\begin{equation}\label{eq:29}
	dN_\mathbf{k}=\sum_{\lambda}dN^{(\lambda)}_\mathbf{k}=
	|\mathbf{k}||\widetilde{\bm{\varphi}}(\mathbf{k})|^2\frac{d^3k}{(2\pi)^3}.
\end{equation}
Expression (\ref{eq:29}) can be also obtained by calculating energy distribution in the wave vector space using the Hamiltonian (\ref{eq:12}). Its average in the coherent state is given by
\begin{equation}\label{eq:30}
	\langle\varphi|\mathcal{H}|\varphi\rangle=\hbar c\int\frac{d^3k}{(2\pi)^3}
	|\mathbf{k}|^2|\widetilde{\bm{\varphi}}(\mathbf{k})|^2, 
\end{equation} 
so that the energy distribution reads
\begin{equation}\label{eq:31}
	dW_\mathbf{k}=\hbar c|\mathbf{k}|^2|\widetilde{\bm{\varphi}}(\mathbf{k})|^2\frac{d^3k}{(2\pi)^3},
\end{equation}
and the photon number distribution, $dN_\mathbf{k}=dW_\mathbf{k}/(\hbar c|\mathbf{k}|)$ agrees with (\ref{eq:29}).

One could also construct the coherent state in standard way using directly the annihilation and creation operators of the photon modes, $|\phi\rangle=\exp(\hat{C})$, where
\begin{equation}\label{eq:311}
	\hat{C}=-\sum_{\lambda}\int d\Gamma_\mathbf{k}a^{(\lambda)}_\mathbf{k}\mathbf{e}^{(\lambda)}_\mathbf{k}
	\cdot\bm{\phi}^*(\mathbf{k})-c.c..
\end{equation}
The profiles $\bm{\varphi}$ and $\bm{\phi}$ in wave vector space are related by
\begin{equation}\label{eq:312}
	\bm{\varphi}(\mathbf{k})=2^{-1/2}i\sum_{\lambda}\mathbf{e}^{(\lambda)}_\mathbf{k}
	[\mathbf{e}^{(\lambda)*}_\mathbf{k}\cdot\bm{\phi(\mathbf{k})}],
\end{equation}
which implies $\bm{\phi}(\mathbf{k})=-2^{1/2}i\bm{\varphi}(\mathbf{k})$.
\section{Space-time structure of the pulse}
To discuss the space structure of the pulse in more details specific form of the coherent state profile has to be chosen. The profile $\bm{\varphi(\mathbf{k})}$ is a divergenceless vector so that  
$\mathbf{k}\cdot\widetilde{\bm{\varphi}}(\mathbf{k})=0$. A simple form of the coherent state profile fulfilling the transversality condition is given by
\begin{equation}\label{eq:32}
	\widetilde{\bm{\varphi}}(\mathbf{k})=\mathbf{k}\times\mathbf{e}\psi(\mathbf{k}),
\end{equation}
where $\mathbf{e}$ is a constant unit vector, which in general may be complex, and $\psi(\mathbf{k})$ is a function of the wave vector. The vector $\mathbf{e}\psi$ is an analog of Fourier transform of positive frequency part of the ``superpotential'' $\mathbf{Z}(\mathbf{r},t)$ introduced in Ref.~\cite{IBB1}. With the use of (\ref{eq:32}) the vector potential (\ref{eq:18}) can be written as
\begin{equation}\label{eq:33}
	\bm{\mathcal{A}}(\mathbf{r},t)=\frac{1}{8\pi^3}\sqrt{\frac{\hbar}{2\varepsilon_0 c}}
	\left[\mathbf{e}\times\nabla f^{(+)}(\mathbf{r},t)+c.c\right],
\end{equation}
where
\begin{equation}\label{eq:34}
	f^{(+)}(\mathbf{r},t)=\int d^3k\psi(\mathbf{k})e^{-ikx}.
\end{equation}
According to (\ref{eq:27}) the normalization condition for the profile of coherent state is
\begin{equation}\label{eq:35}
	\int\frac{d^3k}{(2\pi)^3}|\mathbf{k}||\widetilde{\bm{\varphi}}(\mathbf{k})|^2=N,
\end{equation}
where $N$ is the shorthand notation for the mean number of photons $\langle\hat{N}\rangle_\varphi$. The number of photons in the pulse determines its intensity and the intensity of its electric field as well. In terms of the function $\psi$ the normalization condition reads
\begin{equation}\label{eq:36}
	\int\frac{d^3k}{(2\pi)^3}|\mathbf{k}||\mathbf{k}\times\mathbf{e}|^2|\psi(\mathbf{k})|^2=N.
\end{equation}
The function $\psi(\mathbf{k})$ does not need to be regular for $\mathbf{k}\rightarrow\mathbf{0}$. The only requirement is that that the integral determining $f^{(+)}$ be convergent. Obviously, also the integral in (\ref{eq:36}) must exist, but due to higher positive powers of $|\mathbf{k}|$ in the integrand formula (\ref{eq:34}) imposes stronger limitations on the behavior for small wave vectors. 
The scalar profile $\psi$ will be assumed in the factorized form
\begin{equation}\label{eq:37}
	\psi(\mathbf{k})=Bp(|\mathbf{k}|)\chi(k_z)g(\mathbf{k}_\perp),
\end{equation}
with the function $\chi$, called also the longitudinal profile, strongly peaked at $k_z=k_0$ and the transverse profile $g(\mathbf{k}_\perp)$ significantly narrower than $\chi(k_z)$. The profile of this type corresponds in the configuration space to a pulse moving in the $z$-direction with the electric field oscillating with the central frequency $\omega_0=ck_0$, and with the longitudinal size significantly smaller than the transverse one. The factor $p(|\mathbf{k}|)$ will be assumed in the form $p(\mathbf{k})=|\mathbf{k}|^\beta$, where the power $\beta$ can be negative. It is shown in Appendix B that for the integral in (\ref{eq:34}) to exist this parameter must fulfill the inequality $\beta>-2$. For practical reasons only integer powers will be kept, either $0$ or $-1$. Whereas for a strongly peaked longitudinal profile the choice of $\beta$ does not have a significant impact on the shape of $f^{(+)}$, it is important for the shape of the photon number distribution (\ref{eq:29}), especially at the long wavelength edge. $B$ is the normalization constant to be determined from (\ref{eq:36}). The transverse profile will be assumed in the Gaussian form
\begin{equation}\label{eq:38}
	g(\mathbf{k}_\perp)=\exp\left(-\frac{\mathbf{k}_\perp^2}{2\sigma^2}\right),
\end{equation}
where $\sigma$ is a parameter determining width of the transverse profile; its inverse is of the order of the transverse width of the pulse in the configuration space. Form of the longitudinal profile $\chi(k_z)$ depends on the shape of the pulse envelope. For a Gaussian envelope it is also given by a Gaussian. In the SFA-type calculations the commonly used model of the pulse has the cosine squared-envelope~\cite{Milosevic} which obviously corresponds to a non-Gaussian profile in the wave vector space. If width of the longitudinal profile $\chi$ (e.g. full width at half-maximum) is of the order of $\kappa$ electromagnetic field of the pulse is confined in a region of the volume $\sim (\kappa\sigma^2)^{-1}$ moving with the velocity of light. For a linearly polarized pulse the vector 
$\mathbf{e}$ can be chosen as $\mathbf{e}=(0,1,0)$, and the normalization condition (\ref{eq:36}) takes then the form
\begin{equation}\label{eq:39}
	\int\frac{d^3k}{(2\pi)^3}|\mathbf{k}|(k_z^2+k_\perp^2\cos^2\phi)|\psi(\mathbf{k})|^2=N,
\end{equation}
where $\phi$ is the azimuthal angle in cylidrical coordinates in the $\mathbf{k}$-space. Writing the normalization factor in the form $B=B_1N^{1/2}$ and performing angular integration reduces (\ref{eq:39}) to
\begin{eqnarray}\label{eq:40}
	\nonumber\frac{B_1^2}{8\pi^2}\int_{-\infty}^{\infty}dk_z\int_0^\infty dk_\perp k_\perp (k_z^2+k_\perp^2)^{1/2-n}\\ \times(2k_z^2+k_\perp^2)\chi(k_z)^2\exp\left(-\frac{k_\perp^2}{\sigma^2}\right)=1,
\end{eqnarray}
where $n=-\beta$ will be taken either as $0$ or $1$.
Detailed calculation of the normalization constant is given in Appendix B. Using the formula (\ref{eq:37}) for $\psi(\mathbf{k})$ with transverse profile of the form (\ref{eq:38}), and the longitudinal profile strongly peaked for $k_z=k_0$, one can write the approximate expression for the function $f^{(+)}$
\begin{equation}\label{eq:41}
	f^{(+)}=2\pi k_0^{-n} B_2\sqrt{N}\sigma a_0(t)e^{-\frac{1}{2}\sigma^2 a_0(t)\mathbf{r}_\perp^2}
	\widetilde{\chi}(z-ct),
\end{equation}
where
\begin{subequations}\label{eq:42}
	\begin{equation}\label{eq:42a}
		a_0(t)=\left(1+\frac{ict\sigma^2}{k_0}\right)^{-1},
	\end{equation}
	\begin{equation}\label{eq:42b}
		\widetilde{\chi}(z)=\int_{-\infty}^\infty dk_ze^{ik_zz}\chi(k_z),
	\end{equation}
\end{subequations}
and $B_2=B_1\sigma$. Details of this approximate calculation are given in Appendix \ref{B}, where it is also shown that the factor $B_2$ is finite and non-zero in the plane wave limit $\sigma\rightarrow 0,\,N\rightarrow\infty,\,N\sigma^2=const$. In general, the integration in (\ref{eq:34}) and (\ref{eq:39}) could be done analytically under the assumption that dominant contribution to the integral comes from the region in the $\mathbf{k}$-space for which $|k_z|\gg k_\perp$.

In the standard approach to the theoretical description of interaction with strong pulsed fields the pulse is described by a plane wave with the functional dependence on $z$ and $t$ of the type $f(z-ct)$ (z-dependence is suppressed in the dipole approximation). This formally corresponds to the limit $\sigma\rightarrow 0,\, N\rightarrow\infty$, such that $N\sigma^2=const$. This last condition allows to keep finite the number of photons per unit volume in which the electromagnetic field of the pulse is confined. This volume is of the order of $N\sigma^2\kappa$, where $\kappa$ is the parameter measuring width of the longitudinal profile $\chi(k_z)$. It can be then seen from (\ref{eq:41}) that the expression for $f^{(+)}$ has a well definite plane wave limit. However, it has to be noted that, strictly speaking, the notion of local photon density in space (described by a function of coordinate $\mathbf{r}$) cannot be used ~\cite{IBB3}.

It follows from (\ref{eq:33}) that $f^{(+)}$ must be a dimensionless quantity and, with the choice of the longitudinal profile $\chi(k_z)$ also as a dimensionless function, it follows from (\ref{eq:41}) that the dimension of the factor $B_2$ is $(\textrm{length})^{2-n}$, which agrees with (\ref{eq:40}). Since $\chi(k_z)$ is strongly peaked at $k_z=k_0$ it is convenient to shift the integration variable in (\ref{eq:42b}), $k_z\rightarrow k_z+k_0$ and write the expression for $\widetilde{\chi}$ as
\begin{equation}\label{eq:43}
	\widetilde{\chi}(z)=e^{ik_0z}h(z),
\end{equation}
where
\begin{equation}\label{44}
	h(z)=\int_{-\infty}^\infty dk_ze^{ik_zz}\chi(k_z+k_0).
\end{equation}
For dimensional reasons $h(z)$ can be written as $h(z)=\kappa h_1(z)$ with dimensionless $h_1$, which can be also normalized to unity for $z=0$, i.e. $h_1(0)=1$. E.g. for a pulse with the Gaussian envelope one has
\begin{equation}\label{eq:45}
	\chi(k_z)=\frac{1}{\sqrt{2\pi}}e^{-\frac{(k_z-k_0)^2}{2\kappa^2}},
\end{equation}
and
\begin{equation}\label{eq:46}
	h_1(z)=e^{-\frac{1}{2}\kappa^2z^2}
\end{equation}
Expression (\ref{eq:41}) can now be written as
\begin{equation}\label{eq:47}
	f^{(+)}=C a_0(t)e^{-\frac{1}{2}\sigma^2a_0(t)\mathbf{r}_\perp^2+ik_0(z-ct)}h_1(z-ct),
\end{equation}
where $C=2\pi B_2\sqrt{N}\sigma\kappa k_0^{-n}$. According to (\ref{eq:33}) the vector potential is given by
\begin{equation}\label{eq:48}
	\bm{\mathcal{A}}(\mathbf{r},t)=\frac{1}{8\pi^3}\sqrt{\frac{\hbar}{2\varepsilon_0 c}}
	(\hat{\mathbf{x}}\partial_zf-\hat{\mathbf{z}}\partial_xf),
\end{equation}
\begin{figure*}
\begin{tabular}{cc}
\includegraphics[scale=0.5]{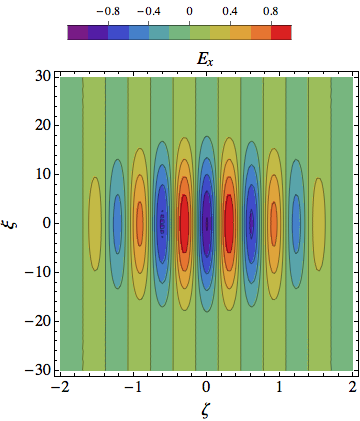}&\includegraphics[scale=0.5]{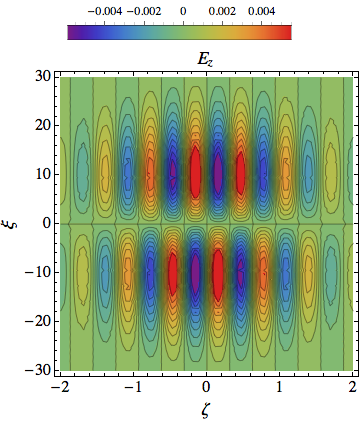}
\end{tabular}
\caption{(Color online) Space images of the transverse $x$-component of the electric field (left panel) and of the longitudinal $z$-component (right panel) at focus, i.e. for $t=0$. The field amplitude $\mathcal{E}_0=1$ a.u., and values of the dimensionless parameters are: $k_0/\kappa=10,\,\sigma/\kappa=0.1.$ The dimensionless coordinates are $(\xi,\eta,\zeta)=\kappa\mathbf{r}$. Note different scales on the 
$\zeta$ and $\xi$ axes. \label{Fig1}}
\end{figure*}
\begin{figure*}
\begin{tabular}{cc}
\includegraphics[scale=0.5]{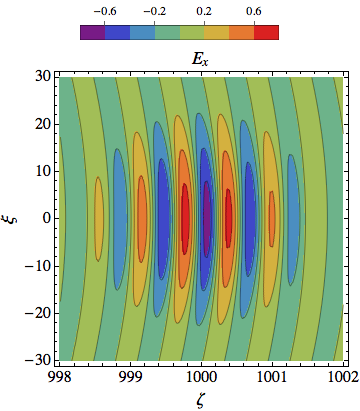}&\includegraphics[scale=0.5]{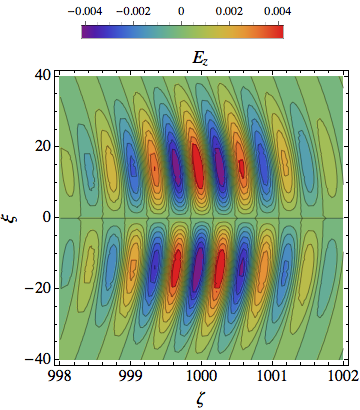}
\end{tabular}
\caption{(Color online) The same as in Fig. \ref{Fig1} but for the dimensionless time $\tau=c\kappa t$ equal to 1000, i.e. for $s^2\tau/u_0=1$. In this case $\zeta=1000$ corresponds to the center of tha pulse and maximum value of $\mathcal{E}_x$.  \label{Fig2}}
\end{figure*}
where $f=f^{(+)}+f^{(+)*}$. It follows from (\ref{eq:47}) that
\begin{widetext}
\begin{equation}\label{eq:49}
	f(\mathbf{r},t)=2C|a_0(t)|h_1(z-ct)e^{-\frac{1}{2}\sigma^2a_r(t)\mathbf{r}_\perp^2}
	\cos[g(\mathbf{r},t)+\gamma(t)],
\end{equation}
where
\begin{equation}\label{eq:50}
	g(\mathbf{r},t)=k_0(z-ct)-\frac{1}{2}\sigma^2a_i(t)\mathbf{r}_\perp^2,
\end{equation}
$\gamma(t)$ is the phase of the function $a_0(t)$ (\ref{eq:42a}), and $a_r$, $a_i$ denote, respectively, the real and imaginary part of $a_0(t)$. Components of the vector potential read
\begin{subequations}\label{eq:51}
	\begin{equation}\label{eq:51a}
		\mathcal{A}_x(\mathbf{r},t)=\frac{1}{4\pi^3}\sqrt{\frac{\hbar}{2\varepsilon_0c}}C
		|a_0(t)|e^{-\frac{1}{2}\sigma^2a_r(t)\mathbf{r}_\perp^2}h_1(z-ct)
		\left\{-k_0\sin[g(\mathbf{r},t+\gamma(t)]+\frac{h'_1(z-ct)}{h_1(z-ct)}\cos[g(\mathbf{r},t+\gamma(t)]\right\},
	\end{equation}
	\begin{equation}\label{eq:51b}
		\mathcal{A}_z(\mathbf{r},t)=\frac{1}{4\pi^3}\sqrt{\frac{\hbar}{2\varepsilon_0c}}C\sigma^2x|a_0(t)|^2
		e^{-\frac{1}{2}\sigma^2a_r(t)\mathbf{r}_\perp^2}h_1(z-ct)\cos[g(\mathbf{r},t)+2\gamma(t)].
	\end{equation}
\end{subequations}
The vector potential and the electric field have non-vanishing longitudinal component, $\mathcal{A}_z$ and $\mathcal{E}_z$, which for short pulses with relatively large transverse spatial extension is small as compared to $\mathcal{A}_x$~\cite{DiChiara}. In the plane wave limit the longitudinal component vanishes and the transverse component of the vector potential has the limiting form
\begin{equation}\label{eq:52}
	\mathcal{A}_x=\frac{1}{2\pi^2}\sqrt{\frac{\hbar}{2\varepsilon_0c}}B_2j^{1/2}\kappa k_0^{-n} h_1(z-ct)
	\left\{-k_0\sin[k_0(z-ct)]+\frac{h_1'(z-ct)}{h_1(z-ct)}\cos[k_0(z-ct)]\right\},
\end{equation}
\end{widetext}
where $j=N\sigma^2$ is a constant quantity in the plane wave limit. Defining
\begin{equation}\label{eq:53}
	\delta(z)=\tan^{-1}\left[\frac{h'_1(z)}{k_0h_1(z)}\right],
\end{equation}
one can write (\ref{eq:52}) as
\begin{eqnarray}\label{eq:54}
	\nonumber \mathcal{A}_x=-\frac{1}{2\pi^2}\sqrt{\frac{\hbar}{2\varepsilon_0c}}\frac{B_2j^{1/2}}{\cos\delta}
	\kappa k_0^{1-n}\\
	\times h_1(z-ct)\sin[k_0(z-ct)-\delta].
\end{eqnarray}
For a Gaussian pulse $h'_1/h_1=-\kappa^2z$. For a typical laser pulse $\kappa\ll k_0$, so that the additional phase in (\ref{eq:54}) can be approximated as $\delta\approx -\kappa^2z/k_0$. Approximating also $\cos\delta$ by unity we get
\begin{eqnarray}\label{eq:55}
	\nonumber\mathcal{A}_x=-\frac{1}{2\pi^2}\sqrt{\frac{\hbar}{2\varepsilon_0c}}B_2j^{1/2}\kappa k_0^{1-n}
	h_1(z-ct)\\
	\times\sin[k_1(z-ct)],
\end{eqnarray} 
where the shifted wave number of the carrier wave is $k_1=k_0+\kappa^2/k_0$, and shifted frequency 
$\omega_1=ck_1=\omega_0+c\kappa^2/k_0$.
In the dipole approximation vector potential of the pulse is usually written as
\begin{equation}\label{eq:56}
	\mathcal{A}_x(t)=\frac{\mathcal{E}_0}{\omega}h(t)\sin(\omega t),
\end{equation}
where $\mathcal{E}_0$ denotes maximum value of the electric field. By comparison of (\ref{eq:55}) and (\ref{eq:56}) we can estimate the electric field amplitude in the present case as
\begin{equation}\label{eq:57}
	\mathcal{E}_0=\frac{1}{2\pi^2}\sqrt{\frac{\hbar}{2\varepsilon_0c}}B_2j^{1/2}\kappa k_0^{1-n}\omega_0,
\end{equation}
with the approximation $\omega_1\approx\omega_0$. The vector potential (\ref{eq:51}) can be therefore written as
\begin{subequations}\label{eq:58}
	\begin{eqnarray}\label{eq:58a}
	\nonumber	\mathcal{A}_x(\mathbf{r},t)=\frac{\mathcal{E}_0}{\omega_0}|a_0(t)|
		e^{-\frac{1}{2}\sigma^2a_r(t)\mathbf{r}_\perp^2}h_1(z-ct)\\
		\times\left[-\sin\lambda(\mathbf{r},t)+\frac{h_1'(z-ct)}{k_0h(z-ct)}\cos\lambda(\mathbf{r},t)\right],
	\end{eqnarray}
	\begin{eqnarray}\label{eq:58b}
		\nonumber\mathcal{A}_z(\mathbf{r},t)=\frac{\mathcal{E}_0}{\omega_0}\frac{\sigma^2}{k_0}x
		|a_0(t)|^2 e^{-\frac{1}{2}\sigma^2a_r(t)\mathbf{r}_\perp^2}\\
		\times h_1(z-ct)\cos[\lambda(\mathbf{r},t)+\gamma(t)],
	\end{eqnarray}
\end{subequations}
where [c.f. Eq.(\ref{eq:50})]
\begin{equation}\label{eq:59}
	\lambda(\mathbf{r},t)=g(\mathbf{r},t)+\gamma(t).
\end{equation}

Electric field is calculated as $\boldsymbol{\mathcal{E}}=-\partial_t\boldsymbol{\mathcal{A}}$. Fig. \ref{Fig1} shows space images of the $x$- and $z$ - components of the electric field of the pulse in the $x-z$ plane at the focus, i.e. for $t=0$. The longitudinal component of the field is much smaller than the transverse one and vanishes along the $x$-axis. These properties are similar to those of the pulse modelled by non-paraxial field ~\cite{DiChiara}. Time dependence of the pulse is described by the functions $|a_0(t)|$, $a_r(t)$, $a_i(t)$ and $\gamma(t)$ in (\ref{eq:58}). In general, transverse dimensions of the pulse increase during time evolution. This can be easily seen from the factor $\exp[{-(1/2)\sigma^2a_r(t)\mathbf{r}_\perp^2}]$ in (\ref{eq:58}), where the function $a_r(t)$ decreases at large $t$ as $t^{-2}$, and therefore width of the Gaussian function grows. With increasing transverse dimensions the density of energy carried by the pulse decreases, and this is accounted for by the overall factor $|a_0(t)|$ in (\ref{eq:58}), which decreases as $t^{-1}$. One can also note that with increasing time edges of the pulse in the transverse direction lag behind the center of the pulse, see Fig. \ref{Fig2}. This effect occurs due character of the time dependence of the function $\lambda(\mathbf{r},t)$ (\ref{eq:59}), and, in particular, to the time dependence of imaginary part of $a_0(t)$, ie. $a_i(t)$, which for times fulfilling $ct\sigma^2/k_0\sim 1$ is close to its maximum value. For the Gaussian pulse (\ref{eq:46}) the square bracket in the expression (\ref{eq:58a}) for the $x$-component of the vector potential can be written as
\begin{equation}\label{eq:60}
	-\left(1+\frac{(\zeta-\tau)^2}{u_0^2}\right)^{1/2}\sin[\phi(\rho,\tau)],
\end{equation}

where
\begin{eqnarray}\label{eq:61}
	\nonumber \phi(\rho,\tau)=u_0(\zeta-\tau)+\frac{1}{2}s^2\rho^2\frac{u_0s^2\tau}{u_0^2+s^4\tau^2}\\ +\tan^{-1}\left(\frac{\zeta-\tau}{u_0}\right)
	-\tan^{-1}\left(\frac{s^2\tau}{u_0}\right).
\end{eqnarray}
The dimensionless variables and parameters are $\rho=\kappa r_\perp,\,\zeta=\kappa z,\,\tau=c\kappa\tau,\,u_0=k_0/\kappa,\,s=\sigma/\kappa$. For $\tau=0$, as in Fig. \ref{Fig1}, the phase $\phi$ does not depend on transverse distance $r_\perp$. Therefore the line of constant phase, where the sine function in (\ref{eq:60}) has constant value, is in the $x-z$ plane a straight line parallel to the $x$-axis. For $\tau>0$ such that $s^2\tau/u_0\sim 1$ the $\rho$ dependence of the phase becomes significant and the values of $\zeta$ for which $\phi(\rho,\tau)=const$ are for $\rho\neq 0$ smaller than for $\rho=0$, which leads to"bending"  of the pulse; this effect is shown in Fig. \ref{Fig2}. For still larger times, such that $ct\sigma^2/k_0\gg 1$, the phase $\phi$ becomes weakly dependent on the transverse distance $r_\perp$. The pulse approaches then the linearly polarized plane wave with large transverse dimension. Maximum value of the electric field becomes smaller and smaller, since energy carried by the pulse is then spread over larger region.

Expectation values of the magnetic field components are given by (\ref{eq:17}), or, explicitly
\begin{equation}\label{eq:62}
	\mathcal{B}_x=\partial_y\mathcal{A}_z,\,\mathcal{B}_y=\partial_z\mathcal{A}_x-\partial_x\mathcal{A}_z,\,\mathcal{B}_z=-\partial_y\mathcal{A}_x.
\end{equation}
It can be seen from (\ref{eq:58}) that the derivatives $\partial_y\mathcal{A}_z,\,\partial_x\mathcal{A}_z,\,\partial_y\mathcal{A}_x$ in (\ref{eq:62}) are small compared to $\partial_z\mathcal{A}_x$. The $x$-component of the vector potential does not contain the small overall factor $\sigma^2$, as well as its derivative with respect to $z$. In the case of three other derivatives either $\mathcal{A}_z$ contains the $\sigma^2$-factor, or it appears as a result of differentiation in $\partial_y\mathcal{A}_x$. For this reason the dominant component of the magnetic field is $\mathcal{B}_y$ with other components smaller by few orders of magnitude. In the plane wave limit ($\sigma\rightarrow 0$ with $N\sigma^2=const$) only transverse components of the fields survive.
\section{Conclusions}
It has been shown explicitly in this paper how a pulse of laser radiation can be described with the use of the coherent state. The most convenient theoretical tool used for this decription are the space dependent annihillation and creation operators~\cite{IBB,Glauber1,Glauber2} instead of the standard operators pertaining to the photon mode approach. The use of space dependent operators allowed for a direct description of space and time dependence of the electromagnetic field of the pulse; this was achieved by calculating expectation values of the corresponding field operators. Another important characteristic of the pulse of radiation is distribution of photon modes, which was found using the connection between the field description and photon mode (wave packet) description of the quantum electromagnetic field~\cite{Smith}. Although in principle the space-time dependence of the pulsed field could be determined by constructing an appropriate solution of classical Maxwell equation (e.g by choosing an appropriate wave packet profile in the plane wave decomposition), quantum attributes of the coherent state are essential for exact determination of the photon mode distribution. The use of coherent state allows also to find the exact relation between maximum value of the electric field intensity and the total number of photons carried by the pulse [c.f. formula (\ref{eq:57})].

Although standard description of interaction with strong laser fields relies on the plane wave and dipole approximation, details of the pulse structure, also in the plane transverse with respect to the direction of propagation, become important for very strong pulses with shorter wavelengths of the carrier wave~\cite{DiChiara}.

\appendix
\section{Derivation of Eq. (\ref{eq:26})}\label{A}
In order to derive Eqs. (\ref{eq:25}) and (\ref{eq:26}) we note that according to (\ref{eq:1a}) and (\ref{eq:20}) space and time dependent annihilation operators $\hat{\mathbf{d}}(\mathbf{r},t)$ can be expressed by annihilation operator of the photon mode as
\begin{equation}\label{eq:A1}
	\hat{\mathbf{d}}(\mathbf{r},t)=i2^{1/2}c^{-1}\sum_{\lambda}\int d\Gamma_{\mathbf{k}}\omega_{\mathbf{k}}
	\hat{a}^{(\lambda)}_\mathbf{k}\mathbf{e}^{(\lambda)}_\mathbf{k}e^{-ikx}.
\end{equation}
Substituting (\ref{eq:24}) into (\ref{eq:A1}) one obtains
\begin{eqnarray}\label{eq:A2}
	\nonumber \int d^3r\boldsymbol{\varphi}^*(\mathbf{r})\cdot\hat{\mathbf{d}}(\mathbf{r})\\
	=\frac{i}{2^{1/2}}\sum_{\alpha,\lambda}\int\frac{d^3k}{(2\pi)^3}
	\widetilde{\boldsymbol{\varphi}}^*(\mathbf{k})\cdot\mathbf{e}^{(\lambda)}_\mathbf{k}f^{(\lambda)}_\alpha
	\hat{b}_\alpha.
\end{eqnarray}
Defining now $A_\alpha$ according to(\ref{eq:26}) we obtain
\begin{equation}\label{eq:A3}
	\int d^3r\boldsymbol{\varphi}^*(\mathbf{r})\cdot\hat{\mathbf{d}}(\mathbf{r})=\sum_{\alpha}A^*_\alpha\hat{b}_\alpha,
\end{equation}
and (\ref{eq:25}) follows from (\ref{eq:6}) and (\ref{eq:A3}).

\section{}\label{B}
\subsection{Calculation of the normalization factor $B$}
The normalization condition (\ref{eq:40}) can be written in the form
\begin{equation}\label{eq:B1}
	\frac{B^2_1}{8\pi^2}\int_{-\infty}^{\infty}dk_z\chi(k_z)^2H(k_z)=1,
\end{equation}
where
\begin{eqnarray}\label{eq:B2}
	\nonumber H(k_z)=\int_0^\infty dk_\perp k_\perp(k_z^2+k_\perp^2)^{(1/2-n)}\\
	\times(2k_z^2+k_\perp^2)e^{-k_\perp^2/\sigma^2}.
\end{eqnarray}
Introducing dimensionless variable of integration, $x=k_\perp^2/\sigma^2$ one gets
\begin{eqnarray}\label{eq:B3}
	\nonumber H(k_z)=\frac{1}{2}\sigma^{5-2n}\int_0^\infty dx\left(x+\frac{2k_z^2}{\sigma^2}\right)\\
	\times\left(x+\frac{k_z^2}{\sigma^2}\right)^{1/2-n}e^{-x}.
\end{eqnarray}
The integration can be done explicitly with the use of ~\cite{nist}
\begin{equation}\label{eq:B4}
	\int_0^\infty dxx^a(x+b)^ce^{-x}=\Gamma(a+1)U(-c,-a-c,b),
\end{equation}
where $U$ is the confluent hypergrometric function of the second type. The parameter $b$ in (\ref{eq:B4}) corresponds to $k_z^2/\sigma^2$. 

Using asymptotic form of the $U$-function for small argument, $U(a,b,z)\sim z^{1-b}$, one can easily show that the function $H(k_z)$ is finite for $k_z\rightarrow 0$ if $n<3/2$. This condition also leads to the photon number distribution (\ref{eq:29}) regular for $|\mathbf{k}|\rightarrow 0$. It will be shown below that the convergence requirement for the integral determining the $f^{(+)}$ (\ref{eq:34}) also imposes the same limitation on $n$.

Since dominant values of $k_z$ in (\ref{eq:B1}) are close to $k_0$, and $k_0/\sigma\gg 1$, one can use asymptotic form of $U$ for large argument, $U(a,b,z)\sim z^{-a}$. Further, $H(k_z)$ is a slowly varying function compared to $\chi(k_z)$ so that, practically, $H(k_z)$ in (\ref{eq:B1}) can be replaced by $H(k_0)$, which for $k_0/\sigma\gg 1$ reads
\begin{equation}\label{eq:B5}
	H(k_0)=\sigma^2 k_0^{1-2n}\left(k_0^2+\frac{1}{2}\sigma^2\right),
\end{equation} 
and (\ref{eq:B1}) can be written as
\begin{equation}\label{eq:B6}
	\frac{B_1^2}{8\pi^2}\sigma^2 k_0^{1-2n}\left(k_0^2+\frac{1}{2}\sigma^2\right)
	\int_{-\infty}^\infty dk_z\chi(k_z)^2=1.
\end{equation}
For the Gaussian longitudinal profile (\ref{eq:45}) one gets
\begin{equation}\label{eq:B7}
	B_2\equiv B_1\sigma=\frac{4\pi^{5/4}}{\kappa^{1/2}k_0^{1/2-n}(k_0^2+\sigma^2/2)^{1/2}}.
\end{equation}
\subsection{Calculation of the function $f^{(+)}$}
Substituting (\ref{eq:37}) into (\ref{eq:34}) and performing angular integration in cylindrical coordinates one gets
\begin{equation}\label{eq:B8}
	f^{(+)}=2\pi B\int_{-\infty}^\infty dk_z\chi(k_z)G(k_z)e^{ik_zz},
\end{equation}
where
\begin{eqnarray}\label{eq:B9}
	\nonumber G(k_z)=\int_0^\infty dk_\perp k_\perp(k_z^2+k_\perp^2)^{-n/2}\\
	\times e^{-ic\sqrt{k_z^2+k_\perp^2}t}
	 e^{-\frac{k_\perp^2}{2\sigma^2}}J_0(k_\perp r_\perp),
\end{eqnarray}
and $J_0$ is the Bessel function of zero order. The function $G(k_z)$ is regular at $k_z=0$ for $n<2$, and, assuming nonnegative and integer values of $n$, the possibilities are $n=0$ and $n=1$. Since the longitudinal profile $\chi(k_z)$ is strongly peaked for $k_z=k_0$ and the integral over $k_\perp$ in (\ref{eq:B9}) is dominated by small values of $k_\perp$ of the order of $\sigma$, the main contribution to the integral comes from the region of $k$-space in which $k_\perp\ll |k_z|\sim k_0$. One can therefore replace the slowly varying factor $(k_z^2+k_\perp^2)^{-n/2}$ by $k_0^{-n}$, and use an expansion in the oscillating exponential function in (\ref{eq:B9}) 
\begin{equation}\label{eq:B10}
	(k_z^2+k_\perp^2)^{1/2}\approx |k_z|+\frac{k_\perp^2}{2|k_z|},
\end{equation}
so that
\begin{eqnarray}\label{eq:B11}
	\nonumber G(k_z)=k_0^{-n}e^{-ic|k_z|t}\int_0^\infty dk_\perp k_\perp\\
	\times\exp\left[-\frac{k_\perp^2}{2\sigma^2a(t,k_z)}\right]J_0(k_\perp r_\perp),
\end{eqnarray}
where
\begin{equation}\label{eq:B12}
	a(t,k_z)=\frac{|k_z|}{|k_z|+ict\sigma^2}.
\end{equation}
The integration can be performed explicitly using the formula
\begin{equation}\label{eq:B13}
	\int_0^\infty dxx\exp(-bx^2)J_0(cx)=\frac{\exp\left(-\frac{c^2}{4b}\right)}{2b},
\end{equation}
which gives
\begin{eqnarray}\label{eq:B14}
	\nonumber G(k_z)=k_0^{-n}\sigma^2a(t,k_z)\\
	\times\exp\left[-ic|k_z|t-\frac{1}{2}\sigma^2a(t,k_z)r_\perp^2\right].
\end{eqnarray}
Substituting (\ref{eq:B14}) into (\ref{eq:B8}) we obtain
\begin{eqnarray}\label{eq:B15}
	\nonumber f^{(+)}=2\pi B\sigma^2k_0^{-n}\int_{-\infty}^{\infty}dk_z\chi(k_z)a(t,k_z)\\
	\times e^{ik_zz-ic|k_z|t}\exp\left[-\frac{\sigma^2a(t,k_z)r_\perp^2}{2}\right].
\end{eqnarray}
Integration over positive $k_z$-axis, when $|k_z|=k_z$, gives the wave propagating in the positive $z$ direction and the integral over negative values of $k_z$ gives rise to a wave propagating in the negative $z$ direction. Since the profile $\chi(k_z)$ is strongly peaked at $k_z=k_0>0$, and rapidly decreases far off $k_z=k_0$, the counter-propagating wave has very small amplitude and can be disregarded. One can therefore put $|k_z|=k_z$ and extend the integration over entire range of $k_z$ from $-\infty$ to $\infty$, since the contribution from negative values of $k_z$ is very small whatsoever. Disregarding further the $k_z$-dependence of the slowly varying function $a(t,k_z)$, ie. replacing $a(t,k_z)\rightarrow a_0(t)=a(t,k_0)$ in the integrand gives expression (\ref{eq:41}) and further (\ref{eq:47}) for the function $f^{(+)}$.


\begin{thebibliography}{99}
    \bibitem{Milosevic} D.B. Milo\v{s}evi\'c, G.G. Paulus, D. Bauer, W. Becker, J. Phys. B: At. Mol. Opt. Phys. \textbf{39} R203 (2006).
    \bibitem{DiChiara} A.D. DiChiara, I. Ghebregziabher, J.M. Waesche, T. Stanev, N. Ekanayake, L.R. Barclay, S.J. Wells, A. Watts, M. Videtto, C.A. Manusco, B.C. Walker, Phys. Rev. A \textbf{81} 043417 (2010).
    \bibitem{Lax} M. Lax, W.H. Louisell, W.B. McKnight, Phys. Rev. A \textbf{11} 1365 (1975).
    \bibitem{Davis} L.W. Davis, Phys. Rev. A \textbf{19} 1177 (1979).
    \bibitem{Huayu} Huayu Hu and Jianmin Yuan, Phys. Rev. A \textbf{78}, 063826 (2008).
    \bibitem{IBB3} I. Bia\l ynicki-Birula, Z. Bia\l ynicka-Birula, Phys. Rev. A \textbf{88} 052103 (2013) 
    \bibitem{Hellwarth} R. W. Hellwarth, P. Nouchi, Phys. Rev. E \textbf{54} 889 (1996).
    \bibitem{Ziolkowski} R. W. Ziolkowski, Phys. Rev. A \textbf{39} 2005 (1989).
    \bibitem{IBB1} I. Bia\l ynicki-Birula, Phys. Rev. Letters \textbf{80} 5247 (1998).
    \bibitem{IBB2} I. Bia\l ynicki-Birula, Z. Bia\l ynicka-Birula, J. Phys. B: At. Mol. Opt. Phys. \textbf{39} S545 (2006).
    \bibitem{IBB} I. Bia\l ynicki-Birula, Z. Bia\l ynicka-Birula, Phys. Rev. A \textbf{79} 032112 (2009).
    \bibitem{Glauber1} R. J. Glauber, Phys. Rev. \textbf{130} 2529 (1963) .
    \bibitem{Glauber2} R. J. Glauber, Phys. Rev. \textbf{131} 2766 (1963).
    \bibitem{Smith} B. J. Smith, M. G. Raymer, New J. Phys. \textbf{9} 414 (2007).
    \bibitem{QED} I. Bia\l ynicki-Birula, Z. Bia\l ynicka-Birula, \textit{Quantum Electrodynamics} (Pergamon, Oxford, 1975), p. 163.
    \bibitem{IBB3} I. Bia\l ynicki-Birula, Z. Bia\l ynicka-Birula, Phys. Rev. A \textbf{86} 022118 (2012). 
    \bibitem{nist} NIST Digital Library of Mathematical Functions, http://dlmf.nist.gov/13.4.4    
\end{thebibliography}
\end{document}